\documentclass[sigconf]{acmart}
\AtBeginDocument{%
  }

\setcopyright{acmlicensed}
\copyrightyear{2025}
\acmYear{2025}
\acmDOI{XXXXXXX.XXXXXXX}
\acmConference[Conference acronym 'XX]{Make sure to enter the correct
  conference title from your rights confirmation email}{June 03--05,
  2025}{Woodstock, NY}
  
\acmISBN{978-1-4503-XXXX-X/2018/06}

\begin{document}


\title{
Investigating Creativity in Humans and Generative AI Through Circles Exercises}

\author{Runlin Duan}
\email{duan92@purdue.edu}
\affiliation{%
  \institution{Purdue University}
  \city{West Lafayette}
  \state{IN}
  \country{USA}
}
\author{Shao-Kang Hsia}
\email{shsia@purdue.edu}
\affiliation{%
  \institution{Purdue University}
  \city{West Lafayette}
  \state{IN}
  \country{USA}
}

\author{Yuzhao Chen}
\email{chen4863@purdue.edu}
\affiliation{%
  \institution{Purdue University}
  \city{West Lafayette}
  \state{IN}
  \country{USA}
}
\author{Yichen Hu}
\email{hu925@purdue.edu}
\affiliation{%
  \institution{Purdue University}
  \city{West Lafayette}
  \state{IN}
  \country{USA}
}
\author{Ming Yin}
\email{mingyin@purdue.edu}
\affiliation{%
  \institution{Purdue University}
  \city{West Lafayette}
  \state{IN}
  \country{USA}
}

\author{Karthik Ramani}
\email{ramani@purdue.edu}
\affiliation{%
  \institution{Purdue University,}
  \city{West Lafayette}
  \state{IN}
  \country{USA}
}

\renewcommand{\shortauthors}{Trovato et al.}

\begin{abstract}
Generative AI (GenAI) is transforming creativity process. 
However, as presented in this paper, GenAI encounters "narrow creativity" barriers. We observe that both humans and GenAI focus on limited subsets of the design space.
We investigate this phenomenon using the "Circles Exercise," a creativity test widely used to examine the creativity of humans.
Quantitative analysis reveals that humans tend to generate familiar, high-frequency ideas, while GenAI produces a larger volume of incremental innovations at a low-cost. However, similar to humans, it struggles to significantly expand creative boundaries.
Moreover, advanced prompting strategies, such as Chain-of-Thought (CoT) prompting, mitigate narrow creativity issues but still fall short of substantially broadening the creative scope of human and GenAI.
These findings underscore both the challenges and opportunities for advancing GenAI-powered human creativity support tools.
\end{abstract}

\begin{teaserfigure}
  \includegraphics[width=\textwidth]{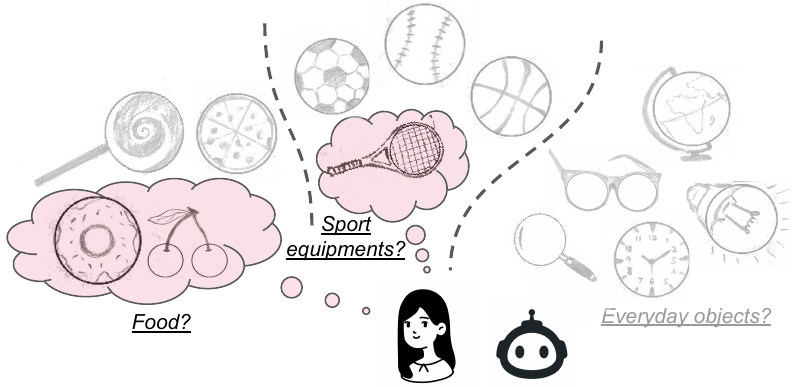}
  \caption{Humans and GenAI tend to explore only a limited subset of the design space during creative tasks.}
  \Description{Teaserfigure}
  \label{fig:teaser}
\end{teaserfigure}


\maketitle
\section{Introduction}

Creativity is the production of ideas or outcomes that are both novel and appropriate to the task or problem \cite{cropley2023intersection}. The creative process involves divergent thinking which requires exploiting existing ideas and exploring unrelated concepts \cite{jirasek2020big, tromp2024creativity}.
However, humans are limited by factors such as intellectual skills, knowledge gaps, and thinking styles \cite{sternberg2006nature}. 

The emergence of Generative AI (GenAI) presents transformative opportunities to explore more creative spaces than humans may without the use of AI, by exposing individuals to vast repositories of knowledge, experiences, and examples \cite{noy2023experimental}. 
Unlike traditional computational methods that rely on predefined datasets \cite{klemmer2002web,chang2012webcrystal,lee2010designing}, GenAI can generate different ideas on demand as references, facilitating the exploration of unconventional and expansive design spaces. 
This unique capability for the potential creative inspirations that GenAI may provide is leading to the use of GenAI in domains such as art \cite{choi2024creativeconnect}, design \cite{liu20233dall}, and writing \cite{chakrabarty2024art,chakrabarty2024creativity}.
In these applications, GenAI acts as a knowledgeable assistant, providing external ideas to enhance human creativity.
However, it remains unexplored to determine the manner in which the human and GenAI perform creative tasks.
Understanding such behaviors is crucial for designing systems that effectively combine human intuition with AI-driven generation to unlock greater potential for human and GenAI to complement each other.
In this paper, our aim is to understand the phenomenon of narrow creativity, defined as the tendency of humans and GenAI to explore only a subset of the available design space. To investigate this phenomenon, we examine: (1) how human narrow creativity is represented in design space exploration, (2) how GenAI exhibits narrow creativity when performing the same tasks, and (3) how advanced prompting methods can be leveraged to enhance GenAI's performance and broaden its creative scope.

We first conduct and quantitatively analyze the human results of a widely used creativity exercise, known as the Circles Exercises \cite{torrance1966torrance,whalley2020paperclips,xiong2024serious}. 
Our analysis provides insights into common aspects of human narrow creativity, including - categories of idea generation, nature of sketching methods, and approaches to material utilization. 
We then conduct a pilot experiment to investigate how GenAI exhibits narrow creativity in the same circle exercise.
We investigated several prompting strategies to uncover the idea generation capabilities and behavior of GenAI. 
The result reveals that GenAI shows similar patterns of narrow creativity as humans do.
Additionally, we observe that GenAI currently excels at reasoning and idea generation, as evidenced in its textual responses, but falls short in generating visual representations.

Our preliminary observations reveal that GenAI, in its current form, has ingrained constraints that lead to narrow creativity in the Circles exercise. 
Additionally, our work inspires further in-depth research to understand narrow creativity in other domains and modalities such as writing, product design, music composition to advance GenAI-powered creativity support tools.

\section{Related Work}

\subsection{Understanding Human Creativity with Design Spaces}

Human cognitive processes and outputs in creative tasks are often structured \cite{ward1994structured}. Specifically, human creativity is shaped by the properties of instructions, provided materials, or existing exemplars. When tasked with a creative problem, humans tend to operate within a design space inherent in the task. Moreover, the design spaces vary depending on the nature of the task. Many previous works have distilled design spaces to better understand human creativity in various domains, such as writing \cite{lee2024design}, storytelling \cite{fan2024storyprompt}, creating virtual reality scenes \cite{neuhaus2024virtual}, and product design \cite{lupetti2024making}. For instance, in a recent study on creative sketching \cite{williford2023exploring}, which is similar to the 28 Circles Test in this work, the authors identified a design space and categorized creations within it.

On the other hand, as highlighted in \cite{finke1996creative,jansson1991design}, humans often exhibit cognitive fixations during creative tasks, generating new ideas by exploiting specific aspects of their conceptual structure. For example, once someone envisions a soccer ball, they are more likely to think of other sports balls rather than animals. As a result, human creativity can become constrained, limiting exploration to a subset of the design space. To address this issue, prior research proposed tools like Mixplorer \cite{kim2022mixplorer}, which represent multiple people's designs with a design space to inform designers about opportunities to explore the design space more broadly and incorporate elements from others' work. 

\subsection{Augmenting Human Creativity with GenAI}

Recently, Generative AI (GenAI) has been increasingly used to facilitate the creative process in the hope that it can enhance human creativity and overcome human limitations by offering fresh perspectives and breaking cognitive fixations in various disciplines, such as design, art and business . In product design, these tools help generate product concepts or prototypes \cite{lu2024generative,duan2024conceptvis,zhang2024protodreamer}, facilitating rapid exploration \cite{kwon2024designer}. In art, GenAI fosters novel creation for visual arts, concept art, music, and literature, as well as video and animation \cite{epstein2023art,liu2024dreamscaping,shi2023understanding}. In business, GenAI aids brainstorming, strategy development, and decision-making through data-driven insights \cite{nguyen2023generative}.

In HCI, many previous works have leveraged GenAI to develop creativity support tools aimed at increasing human creativity \cite{choi2024creativeconnect, wang2024roomdreaming}. However, the creative capacity of GenAI is often overlooked and underexplored. Our work reveals one of its key limitations, narrow creativity. 
Although GenAI is capable of performing incremental exploration on creative tasks within predefined human-provided constraints, it struggles to autonomously generate beyond ordinary or greater radical creativity \cite{cropley2023intersection}. This limitation restricts creative exploration, but simultaneously underscores GenAI's potential to assist human designers in navigating existing design spaces more thoughtfully.

In this paper, we investigate the narrow creativity issues in GenAI by comparing the outputs of humans and GenAI on the Circles Exercise. Through this comparison, our aim is to highlight the limitation of GenAI in creative tasks and stimulate discussion around it. We hope that this work encourages future researchers to develop strategies to enable more effective and efficient human-GenAI collaboration on creativity tasks.

\section{Methodology}

\subsection{Aspects of Human Narrow Creativity}

In this subsection, we aim to explore the aspects of human narrow creativity within the Circles Excercise. 
Specifically, we describe the methodology for analyzing the human outputs of Circles Excercise collected from a classic creativity exercise conducted as a warm up for a graduate-level product design class. 
The task involves asking students to generate as many creative drawings as possible using a provided sheet containing 28 blank circles. By collecting and examining the submissions from the past four semesters, encompassing a total of 3367 drawings made by 224 students, we plan to identify recurring patterns and categorize various aspects of the students’ creative approaches. Below, we outline the specific aspects of creativity we aim to probe and our rationale for investigating each:


\subsubsection{Categories of Drawn Objects}
Understanding the types of objects students tend to draw provides insight into their creative limits and preferences. We classified these objects into categories such as animals, daily object, human figures, nature elements, and vehicles. This classification will help us analyze whether humans gravitate toward familiar, concrete objects or attempt more abstract and imaginative designs. 

\subsubsection{Artistic Expression Techniques}
The way students choose to express their ideas - be they through simple sketches, detailed illustrations, or the use of color - can significantly impact perception of creativity. We will document and analyze these methods to understand the role of artistic style and technical embellishments in creative problem solving.

\subsubsection{Approaches to Material Utilization}
 We plan to categorize these strategies into direct use, personification, abstraction based on circles, and complex compositions, which are derived from our observation and will be explained in the next section. By examining these approaches, we can identify patterns in how students reinterpret the given constraints and adapt the circles to fit their creative visions. 

\subsection{Probing Human Narrow Creativity working with GenAI}

Building on insights into human creative processes, we develop and conduct pilot experiment on GenAI (OpenAI) to further understand how the narrow creativity is present in the GenAI generated results. Specifically, we adopted GenAI to solve the same creativity tasks, the Circles Exercise. 
Recognizing that GenAI’s outputs are heavily influenced by prompting strategies, we initially conducted experiments using naive prompting approaches. Subsequently, we applied the Chain-of-Thought (CoT) prompting technique to further evaluate and compare the performance. The details of the prompting strategies are recorded in the Appendix.

\subsubsection{Naive Prompting}

Naive prompting involves engaging GenAI with minimal input or context, either through zero-shot prompting or few-shot prompting strategies. These approaches allow us to observe the interpretative and generative capabilities under basic interaction scenarios.

In zero-shot prompting, GenAI is provided with a direct instruction or question without any examples or contextual guidance. For the Circles Exercise, this involved asking GenAI to generate solutions or propose design alternatives based solely on the task description. This method reflects how a user might interact with GenAI without prior knowledge of optimal prompting techniques.
In few-shot prompting, the model is given a limited number of examples or context-specific cues before being tasked with generating a response. For this study, we curated a small set of illustrative examples related to the Circles Exercise, aiming to guide GenAI’s response style without providing exhaustive instructions. Few-shot prompting was used to explore how minimal contextualization influences GenAI’s creative output.

\subsubsection{Chain-of-Thought Prompting}

To extend our investigation, we employed Chain-of-Thought (CoT) prompting, a structured approach designed to guide GenAI through step-by-step reasoning. This technique encourages GenAI to articulate intermediate steps and logical processes before arriving at a final solution.

In the context of the Circle Problem, CoT prompting was implemented by crafting prompts that explicitly requested the model to break down tasks into smaller components. These prompts included instructions for exploring alternative ideas, generating intermediate insights, and progressively refining solutions. By structuring the interaction in this manner, CoT prompting aims to simulate a more systematic and reflective creative process, providing a contrast to the unstructured nature of naive prompting.

\section{Result}

\subsection{Quantative Metrices}
In this paper, we use the distribution of drawn object categories and approaches to material utilization to illustrate the narrow creativity of human and GenAI. The object categories were determined based on a coding framework developed from an initial analysis of common design features and functional elements observed in the dataset. Each drawing was manually coded into a category according to its primary function and physical characteristics, following a structured coding process. This process involved two coders who categorized designs independently and resolved disagreements through discussion to ensure consistency and reliability in the coding scheme. The material utilization approaches were categorized by identifying and coding how materials were incorporated into the designs (e.g., direct use of materials and complex composition).

According to research on creativity evaluation \cite{tromp2024creativity, knight2015managing,li2008exploration}, creativity on generating new ideas are best understood by evaluating the exploration and exploitation of design space. 
To investigate the narrow creativity of human and GenAI by analyzing their exploration and exploitation of the explored design space, we developed the following quantitative metrics. 

\begin{itemize}
  \item \textbf{Number of used categories (\# of used cat.)} This metric represents the average number of distinct categories used by each participant during the task.
  \item \textbf{Number of frequent categories (\# of the equation cat.)} To identify the frequently used categories, we calculate the average and standard deviation of the number of circles within each category for an individual. A category is classified as 'frequent' if the number of circles in that category exceeds the average of the individual.
  \item \textbf{Number of highly frequent categories (\# of highly freq. cat.)} This metric identifies categories that are "highly frequent." A category is classified as such if the number of circles it contains exceeds the average by more than one standard deviation.

  \item \textbf{Proportations of drawings in the frequent categories (\% of the frequency category)} This metric counts the proportions of drawings made by an individual within their frequent categories.
  \item \textbf{Proportations of drawings in the highly frequent categories (\% of freq cat.)} Similar to the previous metric, this counts the total number of drawings within an individual’s highly frequent categories.
\end{itemize}

Exploration activity can be measured by the number of categories that an individual has used, while exploitation activity can be assessed by the number of frequent and highly frequent categories that an individual has. An individual who possesses a strong exploration mindset will have a relatively high number of categories used. In addition, the gap between the number of categories used and the number of frequent categories reveals the balance between exploration and exploitation.
When individuals explore multiple categories, a small gap indicates a good balance.

In order to further investigate the inclination between exploration and exploitation, we analyze the distribution of the circles among categories, which reveals how concentratedly an individual focuses on the frequent categories when doing the assignment. If a person has a large portion of circles that fall into the frequent or highly frequent categories, we can conclude that they lean toward exploitation and have narrow creativity issues.
Furthermore, conveying newly generated ideas is a part of the creativity process. In this work, we analyze the categories of artistic expression techniques that humans and GenAI frequently adopt. 

\subsection{Insights into Human Narrow Creativity from the Circles Excercise}

Human creative fixation often reflects a tendency to operate within familiar and constrained boundaries when engaging in creative tasks.
This tendency is evident in the results of the circle test, where individuals demonstrated preferences in object categories, artistic expression, and approaches to material utilization.
By clustering their creative output into key thematic aspects, we gain insight into the strategies humans used to interpret the task. Structurally encoding these strategies further allows us to pinpoint how fixation manifests in human creativity.
Furthermore, analysis of human creativity serves as a valuable baseline for interpreting similar patterns in the results generated by GenAI, enabling a deeper understanding of its capability and limitations.

\subsubsection{Distribution of Drawn Object Categories}

\begin{figure*}
    \includegraphics[width=0.9\textwidth]{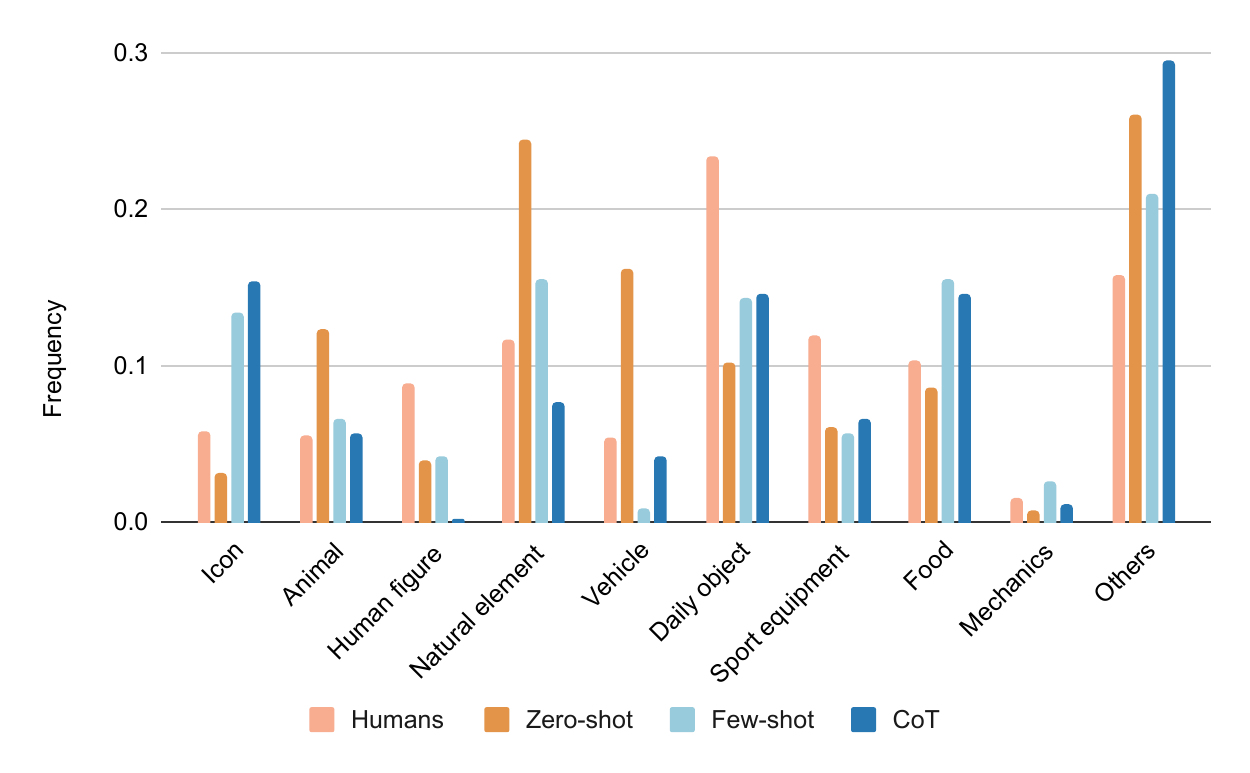} 
    \captionof{figure}{The frequency distribution of categories of objects drawn by humans and GenAI, with a more even distribution across different categories indicating better performance.}
    \label{humanobject}
\end{figure*}

Human creativity tends to cluster around familiar categories of objects. 
Through analysis of students' drawings in the 28 Circles test, we observe that \textbf{daily objects} is the category most frequently used, while the \textbf{ mechanical} is the least adopted category. 
The distribution suggests a tendency to draw inspiration from familiar, easily recognizable elements (daily objects), and reveals a clear preference for relatable, concrete objects over abstract forms or highly imaginative constructs. We provide the frequency of the average number of used categories in Figure \ref{humanobject}.
It shows that in the creativity task, humans exhibit a narrow and skewed bandwidth of creativity, with a significant inclination toward certain categories.

\begin{table*}
  \caption{Quantative analysis of human and GenAI narrow creativity based on the distribution of drawn object categories. For the number of used categories, frequent categories, and highly frequent categories, a higher value indicates better performance. Conversely, for the percentage of objects within frequent and highly frequent categories, a lower value reflects better performance.}
  \label{expression_table_1}
  \begin{tabular}{ccccccccc}
    \toprule
    Object Categories                    &  \multicolumn{2}{c}{Human}          & \multicolumn{2}{c}{Zero-shot} & \multicolumn{2}{c}{Few-shot} &\multicolumn{2}{c}{CoT}\\
    \midrule
                                               & Mean & Std.          & Mean & Std.     & Mean & Std.  & Mean & Std. \\
    \texttt{\# of used cat.}             & 5.6 & 2.1            & 6.6 & 1.6       & 5.6 & 2.0     & 7.5 & 0.97 \\
    \texttt{\# of freq. cat.}            & 2.5 & 1.2            & 2.5 & 0.66       & 2.8 & 1.1     & 3.2 & 0.78 \\
    \texttt{\# of highly freq. cat.}     & 1.4 & 0.96            & 1.4 & 0.65      & 1.2 & 0.28     & 1.6 & 0.69\\ \hline
    \texttt{\% of freq cat.}             & \multicolumn{2}{c}{70}             & \multicolumn{2}{c}{81}        & \multicolumn{2}{c}{77}      & \multicolumn{2}{c}{70}   \\
    \texttt{\% of highly freq. cat. }    & \multicolumn{2}{c}{47}             & \multicolumn{2}{c}{48}        & \multicolumn{2}{c}{43}      &\multicolumn{2}{c}{45}  \\
    \bottomrule
  \end{tabular}
\end{table*}

In this creativity exercise, we observe that human creativity only explores a narrow range when participating in the circle exercise. 
Specifically, an individual likely produces circles concentrated within only a limited number of categories. 
The average number of categories used, frequent categories, and highly frequent categories in all individuals is reported in Table \ref{expression_table_1}. 
Compared with the total number of object categories (10), these results suggest that humans tend to explore a limited subset of categories and frequently narrow their focus even further. 
Humans not only explore narrow perspectives, but also exploit even narrower ones. 
The percentage of objects belonging to frequent categories reveals that the majority of the drawn objects (70\%) fall into frequent categories. 
Thus, the result indicates that the creativity of individuals is strongly inclined towards a limited set of frequent categories.

\begin{figure*}
    \includegraphics[width=0.9\textwidth]{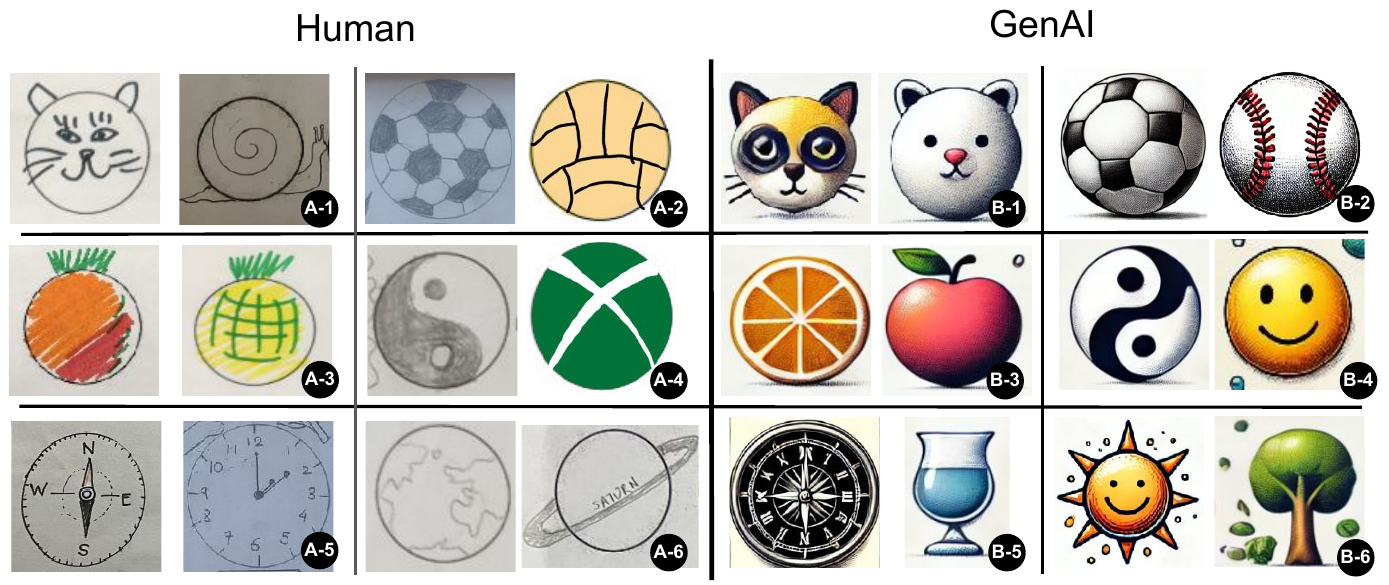} 
    \captionof{figure}{Example of Drawn Object Categories: A) Human Sketched; B) GenAI-Generated, categorized into: 1) Animals, 2) Sport Equipment, 3) Foods, 4) Icons, 5) Daily Objects, and 6) Natural Elements.}
\end{figure*}

\subsubsection{Approaches to Material Utilization}
\begin{figure*}
    \includegraphics[width=0.7\textwidth]{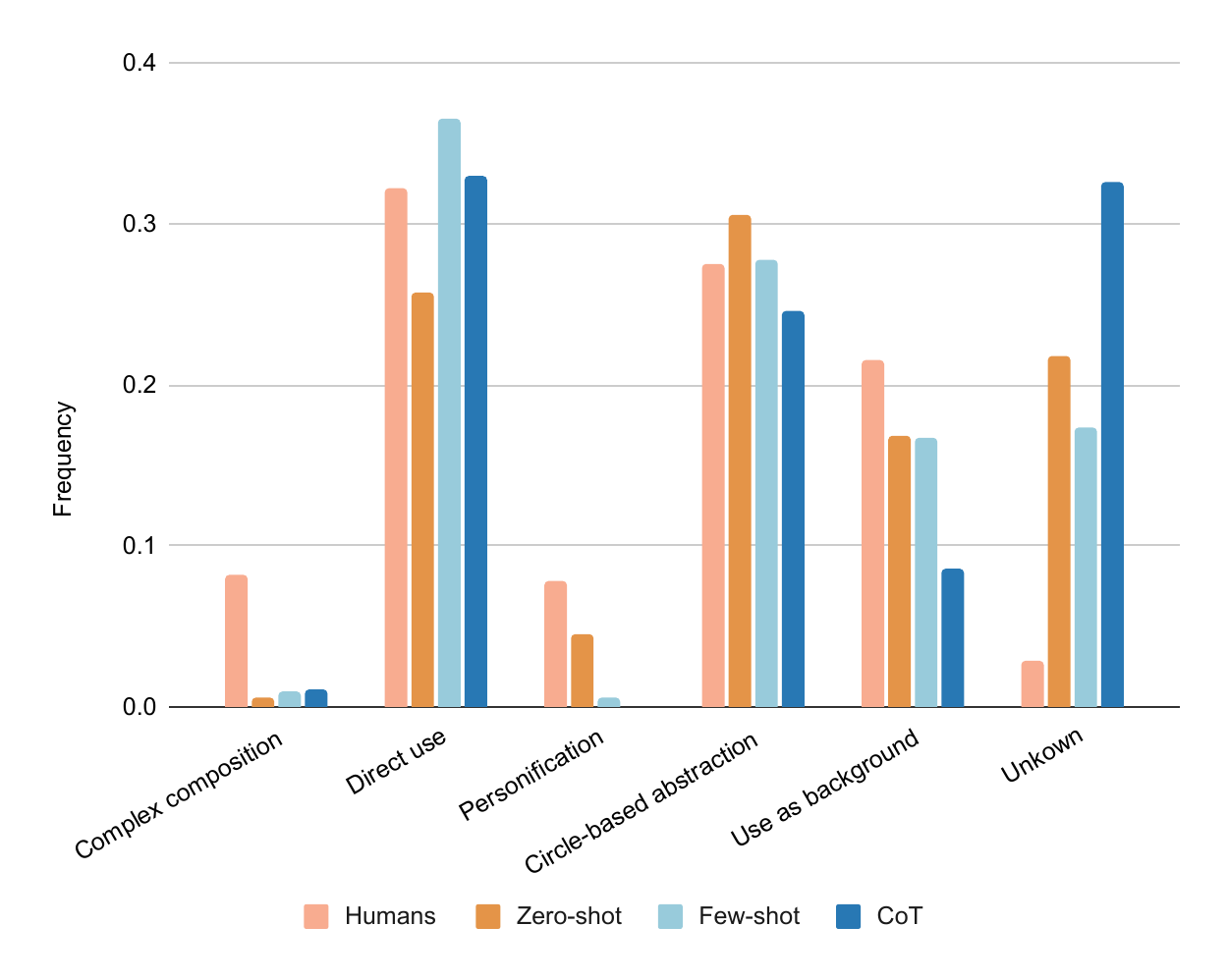} 
    \captionof{figure}{The frequency distribution of approaches to material utilization is analyzed, with a more even distribution across different approaches indicating better performance.}
    \label{humanutilization}
\end{figure*}

Students demonstrate diverse strategies for utilizing the circles for their ideas. These include \textbf{direct use}, where students transform the circles into recognizable objects such as clocks, wheels, or buttons by drawing directly within them; \textbf{personification}, where features of human faces are added to anthropomorphize the circles; \textbf{circle-based abstraction}, where the circles are used as references for similar shapes existing in other objects, such as a tennis racket, lollipop, and gear; \textbf{complex compositions}, where multiple circles were combined to form intricate objects, such as bicycles, ice-cream, and glasses; and \textbf{use as background}, where objects are draw within the circles. The frequency distribution of approaches to material utilization is shown in Figure \ref{humanutilization}. It indicates that humans also present narrow creativity and a skewed preference in terms of material utilization approaches.

\begin{figure*}
    \includegraphics[width=0.9\textwidth]{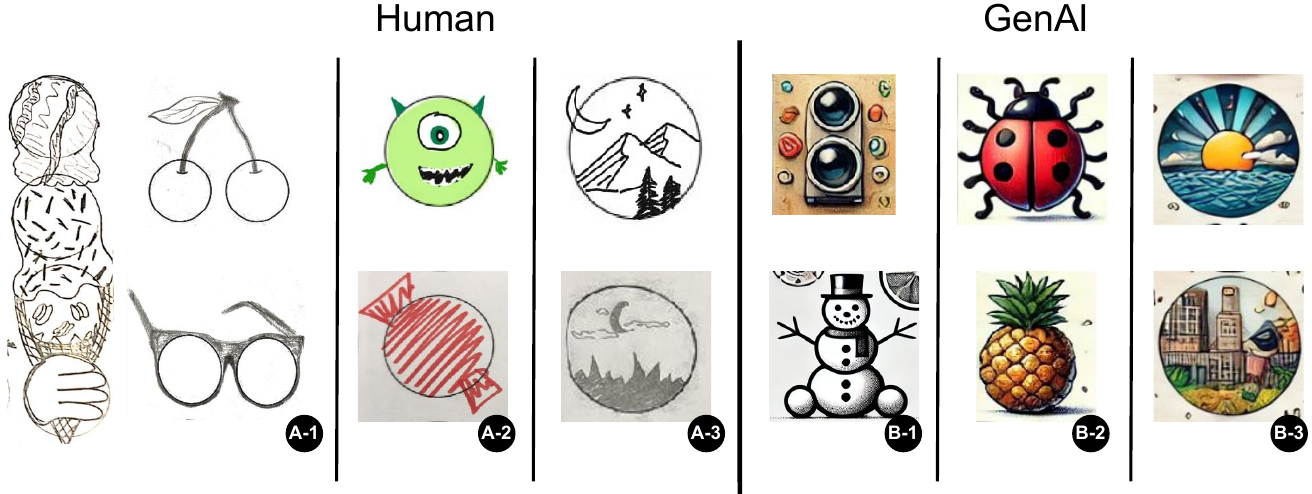} 
    \captionof{figure}{Example of Approaches to Material Utilization: A) Human Sketched; B) GenAI Generated, categorized into: 1) Complex Compositions 2) Circle-based Abstraction 3) Use as Background.}
\end{figure*}


Similarl to the analysis in the above subsection, Table \ref{expression_table_2} provides the average number of approaches used, frequent approaches, and highly frequent approaches. 
Based on the percentage of the frequent approaches, 80\% of the object are proposed based on one frequent approach of using the circles.

\begin{table*}
  \caption{Quantative analysis of human and GenAI narrow creativity based on material utilization approaches. For the number of used approaches, frequent approaches, and highly frequent approaches, a higher value indicates better performance. Conversely, for the percentage of objects within frequent and highly frequent approaches, a lower value reflects better performance.}
  \label{expression_table_2}
  \begin{tabular}{ccccccccc}
    \toprule
    Utilization Approaches                     &  \multicolumn{2}{c}{Human}          & \multicolumn{2}{c}{Zero-shot} & \multicolumn{2}{c}{Few-shot} &\multicolumn{2}{c}{CoT}\\
    \midrule
                                               & Mean & Std.          & Mean & Std.     & Mean & Std.  & Mean & Std. \\
    \texttt{\# of used apch.}                  & 3.0 & 1.2            & 3.5 & 0.96       & 3.4 & 1.2     & 3.6 & 0.69 \\
    \texttt{\# of freq. apch.}                 & 1.5 & 0.61           & 1.9 & 0.86       & 1.4 & 0.50    & 1.7 & 0.67 \\
    \texttt{\# of highly freq. apch.}          & 1.1 & 0.40           & 1.1 & 0.27       & - & -     & - & - \\ \hline
    \texttt{\% of freq apch.}                  & \multicolumn{2}{c}{80}             & \multicolumn{2}{c}{81}        & \multicolumn{2}{c}{63}      & \multicolumn{2}{c}{68}  \\
    \texttt{\% of highly freq. apch. }         & \multicolumn{2}{c}{68}             & \multicolumn{2}{c}{48}        & \multicolumn{2}{c}{51}      & \multicolumn{2}{c}{45}  \\
    \bottomrule
  \end{tabular}
\end{table*}

\subsubsection{Variation in Artistic Expression}

Although students' object categories and artistic expressions exhibit considerable diversity, the artistic styles and techniques they employ show limited variation, reflecting a monotonic and uniform approach to conveying their ideas. The approaches include \textbf{simple sketches}, where many students opt for minimalistic, black-and-white drawings focusing on the core concept; \textbf{detailed illustrations}, with some students enhancing their designs through intricate details that add depth and character. The portion of the two is presented in Table \ref{expression_table_2}, which suggests that humans almost adopt simple sketches. A possible reason might be that humans emphasize task efficiency and believe simple sketches are effective enough to convey their ideas.

\begin{figure*}
    \includegraphics[width=0.9\textwidth]{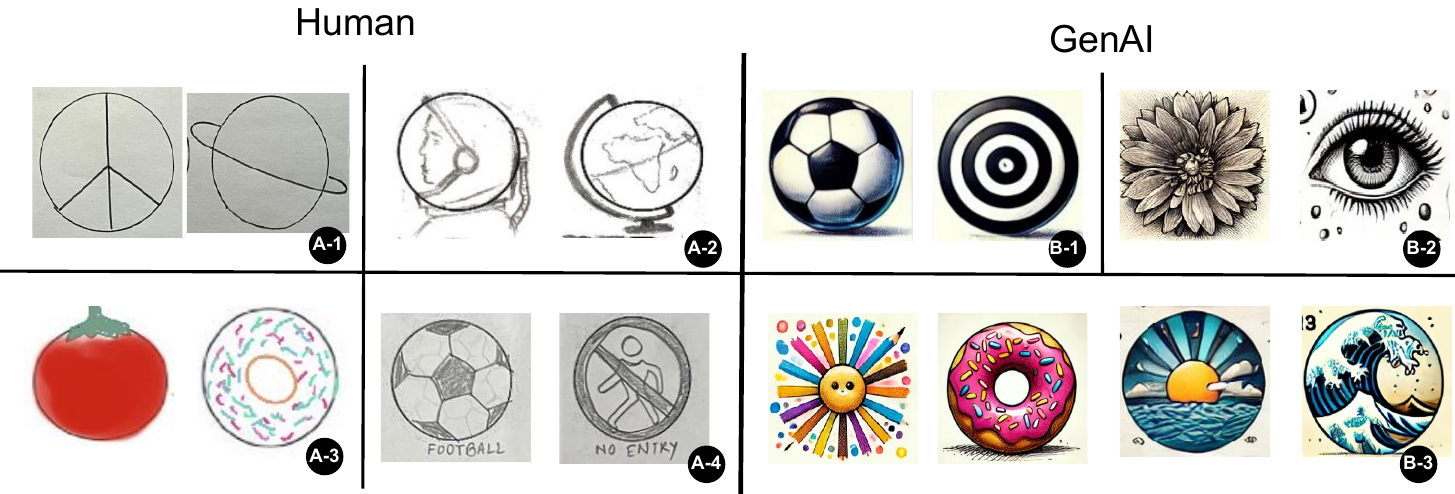} 
    \captionof{figure}{Example of Artistic Expression: A) Human Sketched; B) GenAI-Generated, categorized into: 1) Simple Sketches 2) Detailed Illustrations 3) Use of Color 4) Annotations and Labels. }
    \label{fig: example of art}
\end{figure*}

Besides sketching, some students incorporate additional elements to facilitate their expression. The elements are summarized as follows: \textbf{use of color}, where some students incorporate vibrant colors to enrich their visual representations; and \textbf{annotations and labels}, where some drawings include textual annotation to explain or narrate their drawings, adding an interpretive layer to their visual output. 
The portion of drawings that use the additional elements is reported in Table \ref{expression_table_3}. 
The relatively low portion of drawings with additional elements (5\%) suggests that humans have limited capability to convey their ideas with detailed expressions. It might be because adding additional elements is time-consuming and requires extra resources. 
Figure \ref{fig: example of art} shows that while the GenAI outputs demonstrate enhanced refinement, they require explicit instructions to incorporate annotations or labels.

\begin{table*}
  \caption{The portion of the frequent artistic expression of humans and GenAI.}
  \label{expression_table_3}
  \begin{tabular}{ccccc}
    \toprule
    Artistic Expressions                     & Human          & Zero-shot & Few-shot & CoT\\
    \midrule
    \texttt{\% of simple sketches}           & 95             & 2.9        & 10      & 3.2\\
    \texttt{\% of detailed illustrations}    & 5              & 97       & 89      & 97\\ \hline
    \texttt{\% of use of colors}             & 16             & 52        & 20      & 70\\
    \texttt{\% of use of annotations}        & 13             & 0.0        & 0.0     & 0.0\\
    \bottomrule
  \end{tabular}
\end{table*}

\subsection{Understanding GenAI Narrow Creativity with Different Prompting Strategies}

We present the results of different prompting strategies that are frequently adopted in GenAI-augmented creativity support tools.
We clustered the GenAI result based on the same aspects of narrow creativity as we used for human result.
This analysis offers insights into how AI-augmented creativity can complement or challenge human tendencies, revealing both the limitations and opportunities of prompting strategies in addressing narrow creative issues.

\subsubsection{Naive Prompting}

We conducted a pilot study to evaluate the results of GenAI under the condition of naive prompting.
We adopted the same quantitative metrics that were used to evaluate human creativity.
The zero-shot prompts provided to GenAI were based on a pre-articulated template, as described in the appendix.

For few-shot prompting, we provide each prompt with an examples from the results.

The statistics for zero-shot and few-shot prompting align closely with observations from human data (object categories: human=5.6, zero-shot=6.6, few-shot=5.6; utilization approaches: human=3.0, zero-shot=3.5, few-shot=3.4). 
The distribution of object categories suggests that GenAI, under naive prompting strategies, exhibits a similar pattern of narrow creativity as humans in this task. Interestingly, compared to zero-shot prompting, few-shot prompting produces GenAI results that are more quantitatively aligned with human performance. This observation implies that providing human examples may lead to a similar representation of narrow creativity in GenAI, particularly if the prompts are not further refined or articulated to encourage more diverse outputs.

In terms of artistic expression, humans and GenAI demonstrate differing preferences. Most humans (95\%) prefer to express their ideas through simple sketches. By contrast, GenAI models tend to favor detailed illustrations (zero-shot=97\%, few-shot=89\%), likely due to their stronger image-generation capabilities.
A similar pattern is observed in the use of color. Few-shot prompted GenAI exhibits preferences more aligned with human behavior (human=16\%, few-shot=20\%) due to exposure to human examples during training. This suggests that the inclusion of human examples in few-shot prompting can guide GenAI to mimic certain human tendencies, albeit within the constraints of its learned patterns.

\subsubsection{Chain-of-Thought Prompting}

To better understand the capability of GenAI, we adopted the Chain-of-Thought (CoT) prompting strategy to perform a circle test with GenAI. 
While CoT is considered an advanced technique to enhance GenAI's reasoning capabilities, the experiment results reveal GenAI under CoT prompting demonstrate similiar pattern of narrow creativity as human does.

The result demonstrate that a significant proportion of the objects generated by GenAI under CoT prompting belong to frequently used categories (70\%). This mirrors the behavior seen in other strategies (e.g., zero-shot=81\% and few-shot=70\%), showing that CoT does not substantially expand the variety of generated object categories.
Moreover, 45\% of the objects belong to highly frequent categories, reinforcing the observation that GenAI under CoT also tend to explore on narrowed regions of design space, rather than exploring more diverse ideas.
In terms of material utilization approaches, CoT-generated ideas also exhibit constrained diversity. Approximately 68\% of the approaches employed by GenAI under CoT prompting narrow to the most frequently used methods.
This suggests that CoT does not effectively overcome the bias toward relying on dominant patterns of material utilization.

These results highlight a persistent reliance on frequent object categories and approaches. While CoT improves reasoning capabilities, it does not necessarily enhance the creative breadth of GenAI, as it struggles to generate ideas that break away from the narrow design space.

\section{Discussion}

\subsection{Narrow Creativity Represents in Various Aspects}

Human creativity in structured tasks often relies on familiar and concrete constructs, inherently limiting its scope. 
Recognizing these patterns helps pinpoint the aspects of narrow creativity that constrain human ideation. 
To overcome these barriers, individuals often require external triggers to break beyond habitual thought processes. 
Edward de Bono's concept of lateral thinking is designed to break conventional thinking patterns and overcome the limitations of narrow creativity that constrain human ideation \cite{de1970lateral}. 
This type of understanding and reasoning as well as using structured methods such as SCAMPER \cite{eberle1996scamper} are essential for designing creativity support tools that dismantle such constraints \cite{mymap_ai_idea_generator}. 
By incorporating mechanisms that uncover unexplored design space and facilitate iterative refinement, these tools can significantly broaden creative potential. 
Furthermore, it highlights the need to identify and address narrow creativity specific to diverse domains, such as artwork, product design, and creative writing, to tailor support tools effectively.

\subsection{Constraints of Narrow Creativity in GenAI Outputs}

The analysis of GenAI's outputs demonstrated that, while AI can generate a higher volume of ideas, it is similarly constrained by narrow creativity when not provided with appropriate prompts from human. For example, under zero-shot prompting, GenAI produced generic and repetitive outputs, largely adhering to common categories and lacking the novelty seen in more structured prompts. Few-shot prompting slightly expanded this range, but the AI often mimicked the examples provided, limiting its creative diversity. Even with chain-of-thought prompting, which encouraged incremental refinement, the outputs reflected an iterative rather than groundbreaking approach. These findings suggest that GenAI, like humans, operates within the boundaries of familiar and safe design spaces unless guided to explore further.

\subsection{Develop the Understanding of Narrow Creativity on Generate Creative Tasks}
In this work, we examine the concept of narrow creativity in both humans and GenAI, using the circle exercise as a case study. 
To build on this understanding, we aim to explore narrow creativity in a broader range of generative creative tasks, such as visual arts \cite{choi2024creativeconnect}, writing \cite{lee2024design}, and product design \cite{kwon2024designer}. 
This investigation seeks to uncover general principles and representations of narrow creativity, offering a more comprehensive framework for understanding this phenomenon.
By identifying these principles and representations, we can facilitate ground-breaking idea development and enable both humans and GenAI to navigate the design space using advanced strategies.

\subsection{Advancing Creativity Through Human-GenAI Interaction Mechanisms}

While our study validates the effectiveness of prompting strategies in mitigating GenAI's narrow creativity, it underscores an even greater need for innovative interaction mechanisms between humans and GenAI to foster enhanced creativity.
For instance, incorporating evaluation agents could provide real-time feedback, prompting users to move towards unexplored design space. 
Also, humans could take the lead in identifying and addressing these aspects of narrow creativity, while structured prompts guide GenAI to delve into specific areas. 
This dynamic collaboration allows human users to harness the AI’s extensive generative capacity, directing it toward producing unique and meaningful outcomes. 

\section{Conclusion}
This paper investigated the issue of narrow creativity in both humans and GenAI through the Circles Exercise. 
We began by categorizing and analyzing human drawings to understand the represenation of human narrow creativity. 
Subsequently, we applied the same analysis to GenAI outputs generated using different prompting strategies.
Our findings indicate that both humans and GenAI exhibit narrow creativity, often exploring constrained subsets of the design space. This highlights inherent limitations in the creative processes of both entities.

Our work identifies key challenges and opportunities for advancing GenAI-powered creativity support tools. While sophisticated prompting strategies can partially alleviate narrow creativity, they are insufficient on their own. 
To foster groundbreaking idea development, we suggest future research focus on designing innovative human-GenAI interaction mechanisms and systems. 
Such innovations are crucial for enhancing the efficiency and effectiveness of design space exploration by humans and GenAI.

\bibliographystyle{ACM-Reference-Format}
\bibliography{ref}
\appendix
\section{Appendix}

\subsection{Stardard Circle Exercise Image}
\begin{center}
    \includegraphics[width=0.5\textwidth]{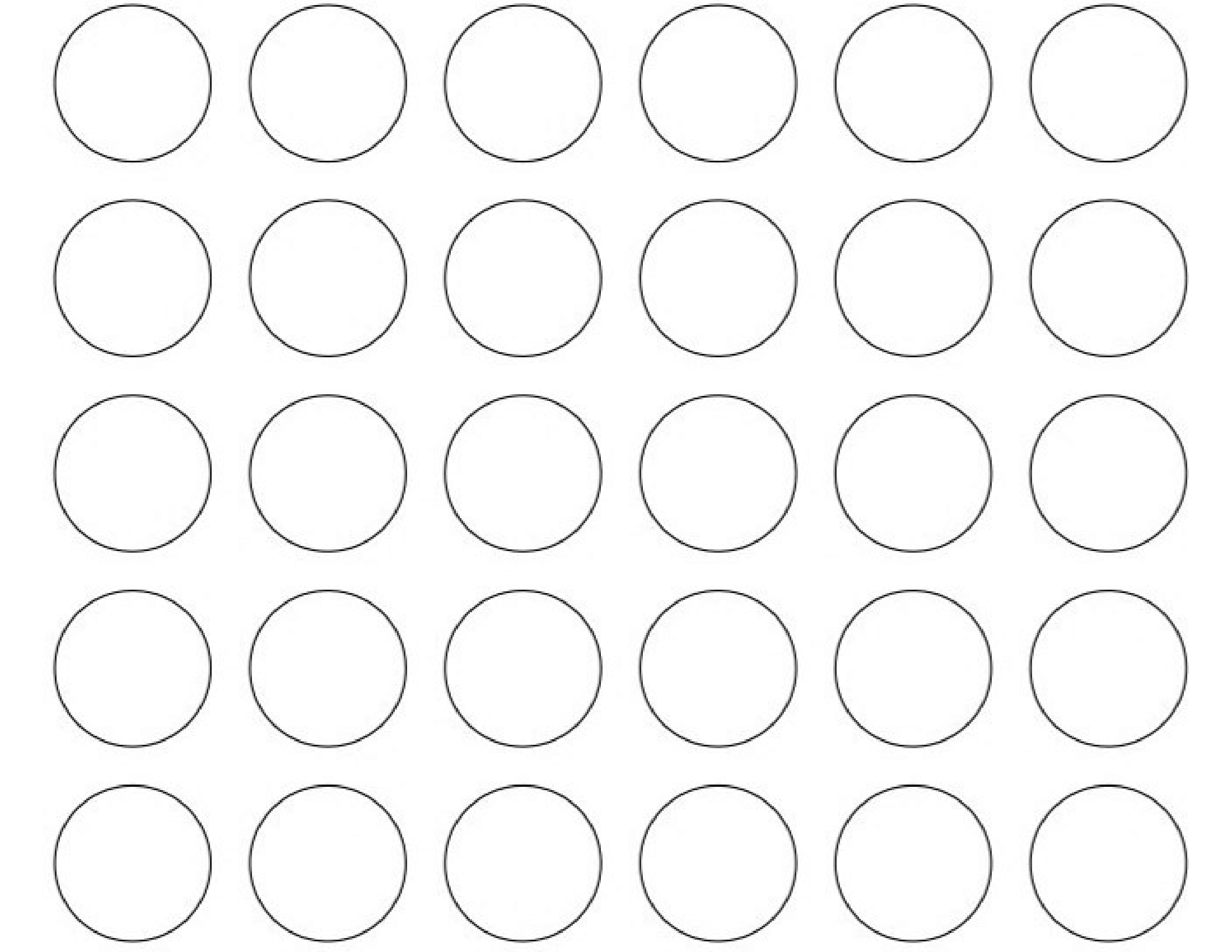} 
    \captionof{figure}{Standard Circle Exercise Image}
\end{center}

\subsection{Circle Exercise Instruction}


"Let's get your creative flow going...Draw as many things as you can - Use each circle in the file attached as the starting point of your creations. Note: The circle should be a part of your creation and 'not just containing the individual sketches inside."

\subsection{Prompt to GenAI}



"Help me complete the creativity challenge, we expect you to generate images of what you have draw. Let's get your creative flow going...Draw as many things as you can in 3 minutes. - Use each circle in the file attached as the starting point of your creations. Note: The circle should be a part of your creation and not just containing the individual sketches inside."

\subsection{Chain of Though Prompt}

\begin{itemize}
    \item Introduce the Challenge Context

“We have a sheet with 30 blank circles. Your task is to create as many different sketches as you can in a short time—ideally 3 minutes or so. Each circle must be used in a way that makes it part of the drawing.”

\item Visualize or Sketch Ideas Quickly

“Quickly brainstorm a wide variety of possible subjects—animals, objects, faces, symbols, doodles—any quick concept you can think of. Remember that each circle should be utilized as an essential element of the final sketch (for example, it might become the center of a flower, the face of a character, the wheel on a vehicle, etc.).”

\item Generate Individual Concepts per Circle

“For each of the 30 circles, propose a short concept description before actually creating the sketch. For example:
Circle 1 → Cartoon bird face
Circle 2 → Bicycle wheel
Circle 3 → Abstract spiral design
… and so on
Brainstorm any simple design that naturally incorporates the circle.”

\item Draw (or Render) Each Idea

“Now, for each circle, create a quick doodle or image that integrates the circle. Make sure the circle remains visible as a key part of the drawing, rather than a container or background only.”

\item Keep It Fast and Fun

“The goal is creativity, so don’t focus too heavily on polished detail. The time constraint is part of the challenge—concentrate on variety and quick execution.”

\item Present the Full Set of Sketches

“Arrange your 30 quick sketches in a grid or sequence that mirrors the original circle layout (5 columns × 6 rows, for instance). Show each circle-based doodle in its own space, with minimal text or labels—let the visuals speak for themselves.”

\item Optional: Reflection/Explanation

“After you’ve completed the sketches, you may add brief captions explaining each idea or the thought process behind it (e.g., ‘Circle turned into a flower with petals around it’). This can help viewers see your creative reasoning.”

\item Encourage Variation and Surprise

“Feel free to let your imagination run wild. Anything from mundane objects (clocks, balloons, and buttons) to fantasy creatures (alien eyes, dragon eggs, planet sketches) is fair game. The more unexpected, the better!”

\item Emphasize the ‘Circle as Part of the Drawing’ Rule

“Throughout every design, ensure the circle remains visibly integrated. Do not hide it entirely behind new elements or treat it merely as a boundary. The circle is part of the shape or composition.”

\end{itemize}

\end{document}